\documentclass{aa}
\usepackage{graphics,psfig,afterpage,multirow,amssymb}

\begin{document}
\title{Crystalline silicate dust around evolved stars\thanks{Based on 
observations with ISO, an ESA project with instruments
funded by ESA Member States (especially the PI countries: France,
Germany, the Netherlands and the United Kingdom) and with the
participation of ISAS and NASA}}

\subtitle{III. A correlations study of crystalline silicate features}

\author{F.J. Molster\inst{1,2,\dagger}
\and L.B.F.M. Waters\inst{1,3}
\and A.G.G.M. Tielens\inst{4,5}
\and C. Koike\inst{6}
\and H. Chihara\inst{7}}

\institute{Astronomical Institute `Anton Pannekoek', University of
Amsterdam, Kruislaan 403, NL-1098 SJ Amsterdam, the Netherlands
\and
School of Materials Science and Engineering,
Georgia Tech, Atlanta, GA 30332-0245, USA
\and
Instituut voor Sterrenkunde, Katholieke Universiteit Leuven, Celestijnenlaan
200B, B-3001 Heverlee, Belgium
\and
SRON, P.O. Box 800, 9700 AV Groningen, The Netherlands
\and
Kapteyn Astronomical Institute, P.O. Box 800, 9700 AV Groningen, The Netherlands
\and
Kyoto Pharmaceutical University, Yamashina, Kyoto 607-8412, Japan
\and
Department of Earth and Space Science, Osaka University, Toyonaka 560-0043,
Japan}

\offprints{F.J. Molster: fmolster@so.estec.esa.nl\\
\mbox{ }$\, \dagger \!\!$ Present address: F.J. Molster, ESA/ESTEC, SCI-SO, Postbus 299, 2200 AG  Noordwijk, The Netherlands}
\date{received date; accepted date}

\titlerunning{A correlations study of crystalline silicate features}
\authorrunning{F.J. Molster et al.}

\abstract{
We have carried out a quantitative trend analysis
of the crystalline silicates observed in the ISO spectra of 
a sample of 14 stars with
different evolutionary backgrounds.
We have modeled the spectra using a simple dust radiative 
transfer model
and have correlated the results with other known parameters.
We confirm the abundance difference of the crystalline silicates in disk and in
outflow sources, as found by Molster et al. (1999a). We found some evidence
that the enstatite 
over forsterite abundance ratio differs, it is slightly higher in the outflow 
sources with respect to the disk sources. It is clear that more data is
required to fully test this hypothesis.
We show that the 69.0 micron feature, attributed to forsterite,
may be a very suitable temperature indicator.
We found that the enstatite is more abundant than forsterite in almost
all sources. 
The temperature of the enstatite grains is about equal to that of the 
forsterite grains in the disk sources but slightly lower in the outflow 
sources.
Crystalline silicates are on average colder than amorphous silicates.
This may be due to the difference in Fe content of both materials.
Finally we find an indication that the
ratio of ortho to clino enstatite, which is about 1:1 in disk sources,
shifts towards ortho enstatite in the high luminosity (outflow) sources.}

\maketitle



\section{Introduction}

The Infrared Space Observatory (ISO) has provided a new and
unprecedented view on the occurence and composition of
circumstellar and interstellar dust. One of the surprises of the
ISO mission was the discovery of ubiquitous crystalline silicates
in circumstellar dust shells of both evolved and young stars (see
e.g. Waters et al. 1996; Waelkens et al. 1996). We have carried
out an extensive study of the presence and properties of
crystalline silicates. The present study is the third in a series,
in which we study these silicates using ISO spectra. In previous
papers (Molster et al. 2001c; 2001d; hereafter Papers I and II respectively) 
we have measured and described the
circumstellar dust features found in the infrared spectra of 17
stars with different evolutionary status.
The majority of these features could be identified
with crystalline olivines (Mg$_{2x}$Fe$_{2-2x}$SiO$_4$) and
pyroxenes (Mg$_{x}$Fe$_{1-x}$SiO$_3$), where $1 \ge x \ge 0$.
J\"{a}ger et al. (1998, hereafter JMD) measured the mass absorption
coefficient of crystalline pyroxenes and olivines  with different
Fe over Mg ratios. Bands of both materials show a shift in the
wavelength position of the peaks to longer wavelengths with
increasing Fe content. The detection of the 69 micron feature,
which is very sensitive to the Fe/Mg ratio (Koike et al. 1993;
JMD), as well as the relative strength of the crystalline silicate
features in the spectrum of IRAS09425-6040 (Molster et al. 1999a), 
led to the conclusion that the crystalline
olivines observed in the ISO spectra are very Mg-rich ($x >
0.95$); the Mg-rich end member of the olivines is called
forsterite. Similarly, the enstatite band at 40.5$\mu$m is sensitive
to the Fe/(Fe+Mg) ratio and points top the presence of Mg-rich pyroxenes.
The identification of the dust species is very important for a
better insight in the formation and evolution of dust. This may
lead to a better understanding of the mass loss process and thus
the evolution of the mass-losing star itself.

There is a clear separation between sources with and
without a dusty disk.
This difference is evident quantitatively in the sense
that the crystalline silicate features are stronger with respect to the
continuum in the sources which are surrounded by a disk (Molster et al. 1999a),
and also qualitatively in the shape of the features, which is
a proof for different dust properties (Paper I).

However, more quantitative statements are necessary to come to a better
understanding of the nature of the circumstellar dust in these objects,
and of their formation and processing history. In order to get these quantitative
statements, a comparison with laboratory measurements is necessary.
Unfortunately the laboratory measurements do not always agree with each other.
In Paper II we discussed different laboratory measurements
of olivines and pyroxenes, and possible causes of discrepancy.
Despite these differences, qualitative agreement with the ISO spectra is
already quite impressive as we will demonstrate in the present study.

In Section~\ref{sec:trends} we discuss trends in the position and width of the
solid state bands.
In Section~\ref{sec9:model} we apply a very simple optically thin dust
model to the spectra to determine a typical
temperature for the dust species.
The results of this modelling are used to look for correlations
which are discussed in Section~\ref{sec:corr}.

Peak positions show variation from source to source.
We adopt the naming in Paper I, which implies that if we refer to
a wavelength position we use
$\mu$m, while if we refer to the name of specific feature we will 
write `micron'.

\section{Peak positions and bandwidths}
\label{sec:trends}

\begin{figure}[t]
\centerline{\psfig{figure=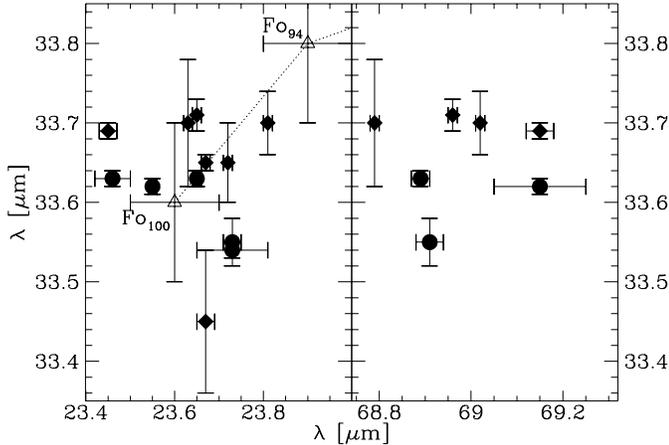,width=88mm,angle=270}}
\caption[]{\small The peak position of the forsterite peak at 23.7 $\mu$m 
versus the peak position of the forsterite peak at 33.6 $\mu$m (left side)
and the forsterite peak position at 69.0 $\mu$m versus
the position at 33.6 $\mu$m (right side). The diamonds denote the disk sources
and the circles denote the outflow sources. The open triangles in the left 
part are room temperature laboratory observations for crystalline 
olivines with 0 and 6\% of [FeO], Mg$_2$SiO$_4$ and 
Mg$_{1.88}$Fe$_{0.12}$SiO$_4$ respectively.
In the right part the laboratory measurements fall off scale due to 
temperature effects (see Fig.~\ref{fig:69T} for these measurements).
The error bars denote 1 $\sigma$ errors in the wavelength position. No obvious 
correlations are visible.}
\label{fig:peak23_33}
\end{figure}

As the measurements in paper~I show, there is a spread in the peak positions 
and bandwidths of the different features (see e.g. Fig.~\ref{fig:peak23_33}, 
where we plotted the spread in the 2 strongest olivine peaks at 23.6 and 
33.6~$\mu$m). There is no clear trend in the spread.
We note that, apart from 89~Her, all disk sources show the 33.6 micron 
features at a longer wavelength than the outflow sources
(see Fig.~\ref{fig:peak23_33}).

Here we will discuss possible causes for this spread. 
One of the best investigated causes for shifts in bands is the chemical 
composition. Difference in the mineralogical composition as well as the 
abundance 
of the different elements are known to change the peak positions. 
Another cause for a change in the peak position is the temperature: a lower 
temperature tends to narrow the features, shift them to shorter wavelengths 
and in some cases even split the band. 
A third method why bands shift and change width is variations in the size and 
shape of the dust grains. 
Changing the crystallinity of the material can also change the appearance of 
the feature. Below we will discuss these four effects more extensively.

\subsection{Composition and temperature}
\label{sec:comp}

\begin{figure*}[t]
\centerline{\psfig{figure=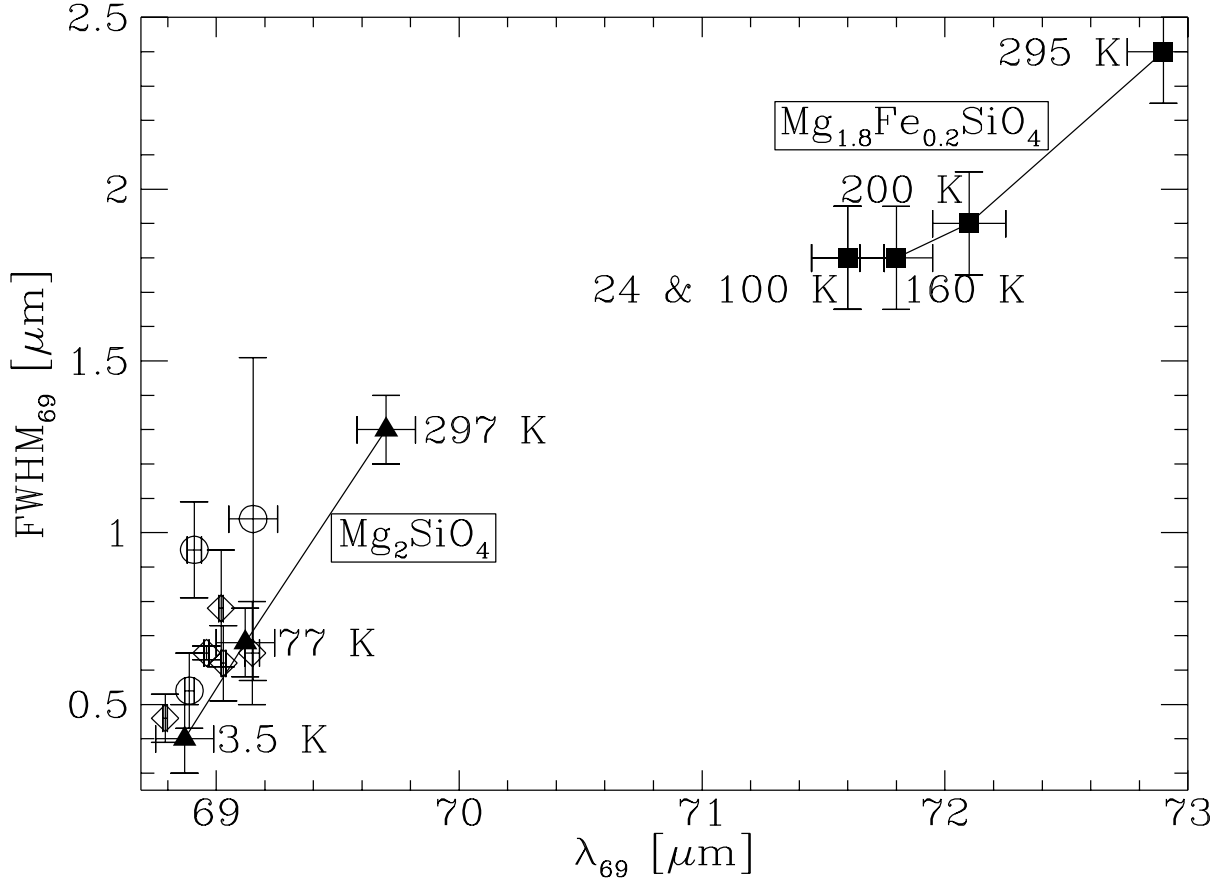,width=160mm,angle=270}}
\caption[]{The observed FWHM and peak wavelength of the 69.0 micron feature
in the spectra of the dust around stars 
(open diamonds for the sources with a disk and open circles for the
sources without a disk) 
and in the laboratory at different temperatures
(filled triangles - forsterite (Fo$_{100}$; 
Bowey et al., 2000), and filled squares - olivine (Fo$_{90}$; 
Mennella et al. 1998)). The temperatures are indicated at each point, and 
within the resolution the 24~K and 100~K for Fo$_{90}$ are similar.
Note that the measurements were not corrected 
for the instrumental FWHM ($\approx 0.29$, for the ISO observations, and
$0.25$ and 1.0$\mu$m for the laboratory observations of respectively 
Fo$_{100}$ and Fo$_{90}$)}
\label{fig:69T}
\end{figure*}

JMD and Koike et al. (1993) showed that the inclusion of Fe in 
(crystalline) silicates will increase the wavelengths of
the peak positions of the different 
features. Since the shift in wavenumbers is rather constant for the different
features and proportional to the [FeO] content (JMD), these shifts are best
seen for the features at the longest wavelengths (see Fig.~\ref{fig:69T}).
However, even a plot of the 69.0 micron feature versus the 33.6 micron 
feature does not show a clear trend (see Fig.~\ref{fig:peak23_33}).
Both Fig.~\ref{fig:peak23_33} and Fig.~\ref{fig:69T} show that the crystalline 
olivines are very Mg-rich and Fe-poor. The wavelength positions are
consistent with
an absence of Fe in these crystals. We conclude that the spread in the 
observations cannot (only) be explained by a very small and changing [FeO] 
content, therefore another mechanism should 
also be responsible.

Another mechanism to shift bands is the temperature.
The 69.0 micron forsterite feature is found in different
laboratory spectra where it always peaks at
69.7 $\mu$m, while in our ISO spectra it is always found around 69.0 $\mu$m.
This shift is significant. Lowering the temperature will shift this 
feature bluewards (e.g. Bowey et al. 2000, Chihara et al. 2001). 
Another effect of a temperature decrease is a narrowing of 
this feature (see Fig.~\ref{fig:69T}).
This is a general property of crystalline features reflecting the
anharmonic interaction of the phonon modes with the thermal phonon bath.
At higher temperature the phonon modes are more excited and their
distribution is broader. Hence, phonon-assisted absorption will shift bands
redwards and will broaden their profiles at higher temperature.
The amount of broadening depends on the origin of the feature.
The 69.0 micron band is one of the best isolated bands in our spectrum to 
test for this effect. This narrowing of the absorption bands with decreasing 
temperature may be responsible for the fact that the observed band widths
(of dust with typical temperatures of $\approx 100$K) in almost all cases is 
smaller than the laboratory widths measured at room temperature.

Besides shifting and narrowing a band, bands can also split in two components
upon lowering the temperature (e.g. Bowey et al. 2000, Chihara et al. 2001).
This might explain why in some
sources we see a blend and in others we find two separate components.
The narrowing (and splitting) of the features might provide an independent
measurement of the temperature of the dust, without knowing anything of
the rest of the spectrum!

We conclude from these comparisons that the crystalline silicates are
very Mg-rich and cold. However, both processes shift the peaks together along
the same line in the wavelength versus FWHM diagram, so other processes must 
play a role too.

\subsection{Shape and size}
\label{sec:shape}

\begin{figure}[t]
\centerline{\psfig{figure=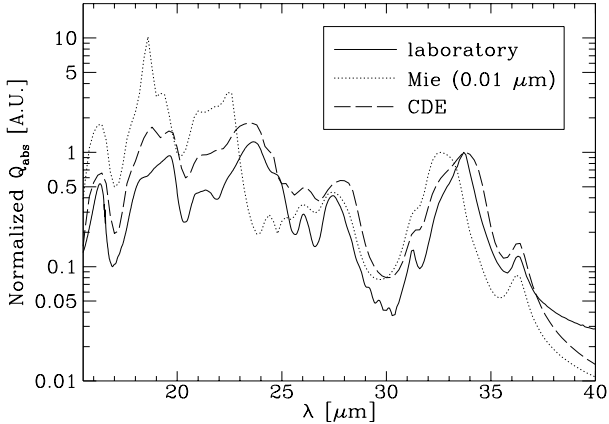,width=80mm,angle=270}}
\caption[]{The absorptivity (Q$_{\rm abs}$) of forsterite measured in the 
laboratory and calculated
from the optical constants of Servoin \& Piriou (1973) for spherical particles
with a radius of 0.01~$\mu$m, using Mie calculations, and for a continuous 
distribution of ellipsoids (having a volume equal to that of a spherical 
particle with a radius of 0.01~$\mu$m). All curves are normalized to the peak
value of the 33.6 micron feature. Note the difference in peak position and
strength between the 2 different shape distributions.
}
\label{fig:shape}
\end{figure}

Strong transitions as found in the crystalline silicates can be very shape 
dependent. Not only the band strength is affected by grain shape, also the
peak position can change dramatically. For instance, the
peak wavelength of the 33.6~micron forsterite feature is located around 
32.7~$\mu$m for spherical grains, while it shifts to 33.8~$\mu$m for a
continous distribution of ellipsoids (see Fig.~\ref{fig:shape}).
The same figure also clearly demonstrates that the relative strength of the 
features is very shape dependent.

Another shape effect, which can play a role in the shape of the spectrum, is 
the preferential growth in the direction of 1 or 2 crystallographic axes
(see e.g. Bradley et al. 1983 and reference in there). 
Although this is not likely to shift and broaden features that much, unless
features corresponding to 2 different axes blend significantly, it 
does play a role in the relative strength of the features.

Difference in the size of the particles can also influence the spectra. 
However, this only becomes important when the size of the particles 
is comparable to the wavelength of the feature. Relatively large 
particles will broaden the feature and shift it to longer wavelengths. 

Finally, we would like to note that coagulation and porousity can also have 
an effect on the width of the bands. This has been theoretically investigated 
for amorphous quartz spheres by Bohren and Huffman (1983) and in the 
laboratory by Koike and Shibai (1994). This effect is difficult to 
quantify, but is likely to play a role for the crystalline silicates.

To conclude, we can say that shape effects might play a role in the spread of 
the peak positions. 

\subsection{Crystallinity}
\label{sec:cryst}

A final consideration should be the crystallinity, i.e. degree of lattice 
order or number density of lattice defects. 
Single crystals without defects show much sharper peaks than
those with some defects. Therefore the width of the peaks might be an 
indication of the crystallinity of the dust grains.

\begin{figure}[t]
\centerline{\psfig{figure=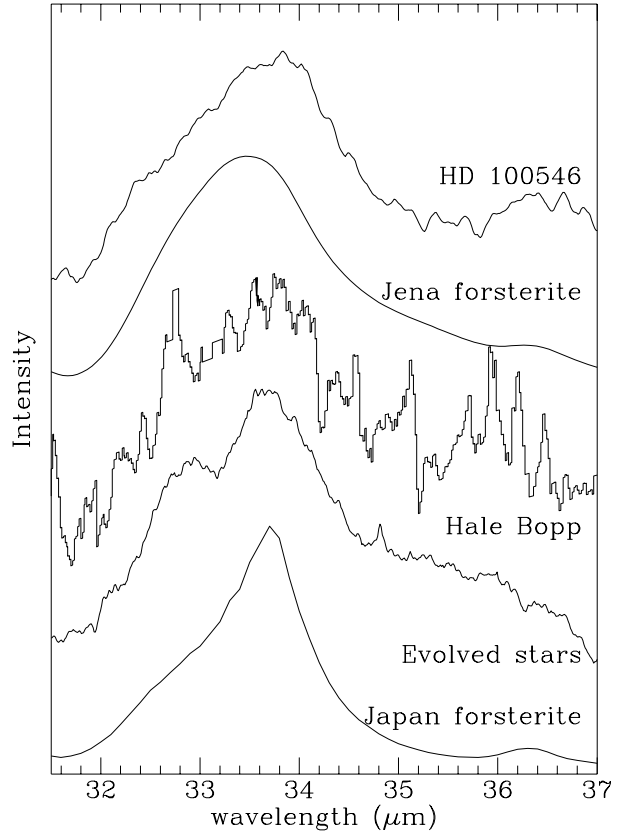,width=80mm,angle=0}}
\caption[]{The 33 micron complex of the young star HD100546 
(Malfait et al. 1998), the comet Hale Bopp (Crovisier et al. 1997)
and an average of 33 micron complexes of evolved stars with evidence 
for a disk (Molster et al. submitted to A\&A), 
together with 2 laboratory measurements of forsterite 
one done by JMD and one by Koike et al. (2000b).
Note the difference in width of the 33.6~micron feature.
}
\label{fig:33fors}
\end{figure}

Fig.~\ref{fig:33fors} shows a comparison between forsterite in different 
environments and 2 laboratory spectra. The forsterite measured by JMD was 
formed from a melt and probably polycrystalline, while the forsterite 
measured by Koike et al. (2000b) was a single crystal. The poly-crystalline
forsterite is expected to have defects where the different crystals
meet each other.

The comparison with the astrophysical spectra suggests that around evolved 
stars nature produces nice single crystals, while around young stars
poly-crystalline material seems to be formed. Around young stars it is 
expected that crystalline silicates form at the inner edge of accretion disks
where the dust particles will (partially) melt. When mixed with cooler 
regions the liquid drops will likely solidify again and crystallize.
In the outflows of evolved stars, gas will 
slowly cool and condense into dust grains. The time and temperature in these
environments are likely to be sufficient to
completely crystallize and get rid of all defects.
It seems that this is not the case for young stars, where the (partially) 
melted dust grains cool more rapidly, being unable to remove all the defects.

Fig.~\ref{fig:33fors} already shows that the crystallinity might shift the
peak position of the feature a few tenth of a micron. So it is likely to play 
a role in the spread.

\subsection{Summary}
\label{sec:sum}

Above we discussed several mechanisms to shift and broaden features. We can 
conclude that the crystalline silicates are Fe-poor and cold. However, the 
reason for the spread in peak position is not well known. 
Shape effects and crystallinity, due to the different environments 
in which these particles are formed, are likely to play a role, but 
a combination with temperature and composition effects cannot be excluded.

Finally, it should be noted that all the features, of which the peak positions 
are plotted in Fig.~\ref{fig:peak23_33} are part of a complex. 
Contributions of nearby
features from other materials, which were not detected as separate features, 
might result in (apparent) differences of the
peak positions. In this respect it is interesting to note that the
33.6~micron feature is the dominant feature in the 33 micron complex and also
seems to show the least spread in the observations. The 23.6 and 
69.0~micron features are part of the 23 and 60 micron complexes, 
respectively, and are much less dominant in these complexes. So relatively 
minor contributions will have less impact on the 33.6 micron feature than
on the 23.6 and 69.0 micron features.

\subsection{Other claimed trends}

In a very limited sample Voors (1999) found a constant separation 
(in $\mu$m) between the 30.6 and the 32.8 micron band, 
which suggest a common species for these two features.
We could not confirm this, but found a relation 
with $\lambda_{x}$ indicating the wavelength of the $x$ micron feature.
Although we found a trend for those two features, we consider it
unlikely that the 2 dust features come from the same material, since their
strength normalized to the 33.6 micron feature does not correlate.
Although we did not further investigate the correlation found, it may reflect 
a common cause, e.g. a difference in general temperature of the dust particles.
 
Based on a study of the crystalline silicate features in AFGL4106,
Molster et al. (1999b) concluded that the forsterite features
were broader than the enstatite features. We have tested
whether this conclusion holds for our larger sample, using the mean results 
from paper~II, and could not confirm their claim. A similar negative result 
was found, when we compared the different laboratory datasets.

\section{The modelling}
\label{sec9:model}

\begin{table*}[ht]
\caption[]{\small The derived temperatures for forsterite $T_{\rm f}$,
enstatite $T_{\rm e}$ and the amorphous silicates $T_{\rm a}$
from our model fits. Also the forsterite to enstatite
ratio is given. The `-' denotes
that it was not possible to derive a realistic value.
$\dagger$ indicates that it is not really possible to fit the spectrum 
with a single temperature, The temperatures here are found by a fit to the 
short wavelength side of the spectrum.
The typical error in the temperature of the crystalline silicates
is 10~K, for the continuum temperature about 20~K, and in the mass
ratio a factor 2. (P)PN = (proto-)planetary nebula, RSG = red supergiant.
}
{\small
\centerline{
\begin{tabular}{|l|l|c|c|c|c|}
\hline
{\multirow{2}{10mm}{Star}}&
{\multirow{2}{10mm}{Type}} &
{\multirow{2}{7mm}{$T_{\rm f}$}} &
{\multirow{2}{7mm}{$T_{\rm e}$}} &
{\multirow{2}{7mm}{$T_{\rm a}$}} &
{\multirow{2}{7mm}{$\frac{{\rm M}_{\rm e}}{{\rm M}_{\rm f}}$}} \\
                &                       &       &       &       &      \\
\hline
\multicolumn{6}{|c|}{\em Disk sources}  \\
\hline
IRAS09425-6040  & C-star with O-dust    & 85    & 100   & 145   & 1.2  \\
NGC6537     & hot PN        & 75    & 65    & 80$^\dagger$   & 5.8    \\
NGC6302     & hot PN        & 65    & 70    & 80$^\dagger$   & 1.0    \\
MWC922      & PPN?, Herbig star?    & 90    & 100   & 140   & 2.7  \\
AC Her      & RV Tauri star     & 100   & 90    & 225$^\dagger$   & 5.0 \\
HD45677     & Herbig star?      & 140   & 140   & 235$^\dagger$   & 5.3 \\
89 Her      & PPN           & 110   & 100   & 320$^\dagger$   & 5.0     \\
MWC300      & PPN?          & 90    & 90?   & 145   & 1.4?    \\
HD44179     & C-PPN with O-dust & 135   & 135   & 120   & 4.0 \\
\hline
\multicolumn{6}{|c|}{\em Outflow sources}   \\
\hline
HD161796    & PPN           & 105   & 80    & 100   & 11.4  \\
HD179821    & post-RSG?     & 75    & 65    & 90    & 3.3   \\
AFGL4106    & post-RSG      & 100   & 80    & 120   & 8.0   \\
NML Cyg     & RSG           & 150   & -     & 180   & -  \\
IRC+10420   & post-RSG      & 90    & -     & 160   & -  \\
\hline
\end{tabular}}}
\label{tab:T_fit}
\end{table*}

In this section we will apply a simple dust emission model
to derive the typical temperature and abundance ratio of
forsterite and enstatite in the dust shells. We assume that the
dust shell is optically thin at infrared wavelengths (which is
reasonable because we see the dust features in emission), the grain size
distribution of the different dust species is similar to what has
been measured in the laboratory ($\ll 2\mu$m, which also is quite realistic,
because the width of the features indicates grains smaller than the wavelength)
and that all grains of a given
composition have the same (single) temperature (this is probably less
realistic, but we only want to get a typical temperature for the dust species
and are at the moment not interested in the temperature distribution).
A comparison of several
laboratory data sets with the observations indicates that the
laboratory data of Koike et al. (2000b) give a good qualitative
match to the observations. We will use this
data set for our modelling. 
We determined an eye-ball spline-fit continuum, maximizing the
continuum and still be smooth (no sudden changes in the slope), both in
$F_{\nu}$ and $F_{\lambda}$. This continuum was derived in a similar way 
(using similar wavelength positions as continuum) for both the stellar 
and the laboratory spectra. Whenever possible we tried to
use the whole wavelength range (SWS + LWS) to determine the
placement of the continuum for the stellar spectra (see also Paper I).
We have fitted the continuum
subtracted spectra with the continuum subtracted forsterite and
enstatite (50\% ortho-enstatite and 50\% clino-enstatite) mass
absorption coefficients multiplied by blackbody functions.

\begin{equation}
F(\nu)_{\rm model} = \sum_{i} B(T_i,\nu) * \kappa(\nu)_{i} * M_{i}
\label{eq:model}
\end{equation}
$F(\nu)_{\rm model}$ is the calculated model flux, $B(T_i,\nu)$
is the blackbody function at temperature $T_i$ of dust species $i$,
$\kappa(\nu)_{i}$ is the mass absorption coefficient of dust species $i$ and
$M_i$ is a multiplication factor which is related to the total
mass of dust species $i$.

The temperature of the blackbody is not necessarily the same for
enstatite and forsterite. In determining the best fit, we varied
the temperature in steps of 5~K. The resulting spectra were
separately scaled to fit the spectrum. This scaling factor is
related to the mass of the dust species. The absolute masses
requires knowledge of the distances to the stars but, for each source the 
masses of the different dust components can be directly compared.
The mineral
mass ratios determined in this paper assume that they have the
same grain size and shape distribution (both around stars and in
the laboratory samples). The best fits were determined by eye and
no $\chi^2$ method has been applied. This method is of sufficient
accuracy given the current quality of the lab data and given the
fact that several prominent dust features still lack
identification, thus strongly affecting any $\chi^2$ method. We found
that the temperature and mass for forsterite could
be determined using the 23 and 33 micron complexes, while
the enstatite values are mainly based on the 28
and 40 micron features.

\begin{figure}[t]
\centerline{\psfig{figure=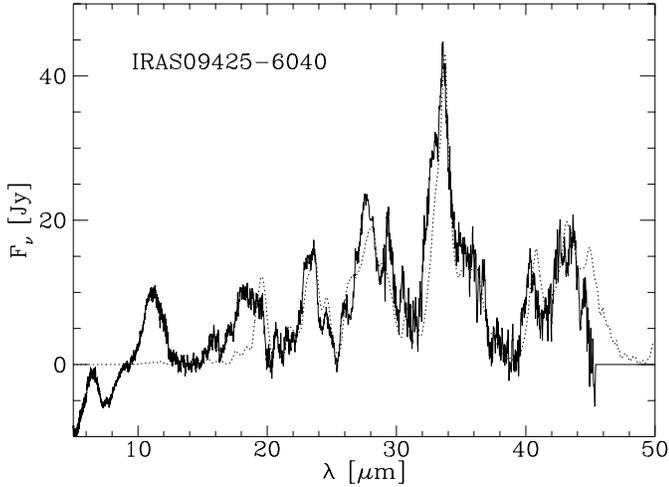,width=88mm,angle=270}}
\caption[]{\small A fit (dotted line) to the continuum subtracted spectrum (solid line) of IRAS09425-6040. $T_f = 85$~K and $T_e = 100$~K.}
\label{fig:iras09_fit}
\end{figure}

The results of this simple fitting procedure are shown in
Fig.~\ref{fig:iras09_fit} to~\ref{fig:afgl4106_fit} and the
derived temperatures and mass ratios are given in
Table~\ref{tab:T_fit}. We also derived an estimate for the
typical temperature of the underlying continuum. For this we assumed 
that the continuum is caused by
small grains with optical constants based on the
amorphous silicate set 1 of Ossenkopf et al. (1992) and a continuous 
distribution of ellipsoids as shape distribution. We fitted
the continuum to the original, not the continuum subtracted,
spectra. An independent fit based on a $Q(\lambda) \sim
\lambda^{-1}$ emissivity law gave similar temperatures. This gave
us confidence that the continuum temperature is reasonably well 
determined in this way. It should be noted that other shape distributions
(e.g. spheres) and other sets of optical constants of amorphous olivines can
easily change the derived temperature by $\pm$ 20~K, more often to higher than 
to lower temperatures. From these fits we could in principle derive a relative 
mass, like in the case of enstatite and forsterite. Although the 
uncertainties in the (mass) absorption coefficients (due to shape, size and 
compositional differences) are systematic, the spread in values makes it
very difficult to interpret them and to compare them with other observations. 
Therefore, we have not given an amorphous over crystalline silicate mass ratio.
However, since the differences between the different datasets are systematic, 
trends can still be derived from these numbers.
For the remainder of this paper we will take the
temperature derived by the fit with the Ossenkopf data set as the
continuum temperature (Table~\ref{tab:T_fit}), because these fits tend 
to produce the best fits. We compared the temperatures
found in this study with those found by Molster et al. (1999b; 2001a), 
and we found a reasonable agreement. Difference in the
temperatures found could often be described to the use of
different laboratory data sets.

Our simple model, consisting of only two crystalline dust
components and a single temperature for each dust component, 
fits most stars very well, see e.g. MWC922
(Fig.~\ref{fig:mwc922_fit}). Still, it is clear that this simple
model is not sufficient to explain all the features. The main
discrepancies between our model fits and the ISO data lie at the
wavelengths below 20~$\mu$m. We note that the three stars with a
continuum temperature above 200~K all show crystalline silicates
in emission in the 10 micron region. The temperature of the
crystalline silicates has been determined based on bands at
wavelengths longwards of 20 micron. These bands are dominated by cool
dust, and the derived low temperatures (Table~\ref{tab:T_fit})
are too low to explain the strength of the crystalline silicate
bands in the 10 micron complex. A second, much warmer, component
must be introduced to explain these 10 micron bands. Likely a
temperature gradient is present in these sources. The discrepancies
shortwards of $20 \mu$m do not solely reflect the presence
of a temperature gradient in these sources, but indicate
that still other dust components must be present. The 18
micron complex is badly fitted. The modelled 19.5 micron feature
(originating from both forsterite and enstatite) is often much
too strong and the modelled 18.0 and 18.9 micron features are
often too weak when compared to the ISO spectra. The too strong
19.5 micron feature might be a radiative transfer effect, since
this feature is less of a problem in the full radiative transfer
modelling (see e.g. Molster et al. 1999b; 2001a). This might indicate
that our assumption of $\tau \ll 1$ is not correct at wavelengths
around 19.5~$\mu$m. The poor fit of the 18.0 and 18.9 micron
features suggests the presence of another dust component.

There is more evidence for the presence of an extra dust
component. The 29.6 and 30.6 micron features also need extra
emissivity, as is very clear in the spectra of NGC6537
(Fig.~\ref{fig:ngc6537_fit}) and of NGC6302
(Fig.~\ref{fig:ngc6302_fit}). In these two sources the 40.5
micron feature is not well fitted, suggesting that the same dust
component which is responsible for the 29.6 and 30.6 micron
features also has a peak around 40.5 $\mu$m.
A possible candidate for this extra dust component is diopside
(MgCaSi$_2$O$_6$), which peaks at the required wavelengths.
However this material also produces strong peaks at other wavelengths,
e.g. at 20.6, 25.1 and 32.1 $\mu$m, which are observed in the ISO data, but 
often not as strong as expected.
Therefore, the identification of the carrier of the 29.6 and 30.6 micron
features remains open. It should be noted, that the temperature and 
relative mass of enstatite are estimated from the 28 and 40 micron 
complexes. Therefore a significant contribution of an unknown dust 
component to one (or both) of these 2 complexes can change the estimated 
temperature and abundance of enstatite.

The 33.0 micron feature is not well fitted, but this feature is
likely to be influenced by instrumental behaviour (see Paper II).
In the 35 micron plateau we clearly miss intensity around 34.8
$\mu$m in all sources. The predicted 69.0 micron feature is often
too weak with respect to the ISO spectra (see e.g.
Fig.~\ref{fig:hd179821_fit}). This may be an indication for the
presence of colder dust, and thus for a temperature gradient, we
will come back to this later.

Apart from all the features that are missing, we also have a problem with
too much intensity predicted by our modelling around 27~$\mu$m.
This excess is mainly due to enstatite, but also forsterite contributes
slightly. We are still looking for an explanation of this phenomenon.

Finally, we did not attempt to fit the absorption profiles. As stated
above we assumed the dust was optically thin.
Also, no attempt was done to fit
the carbon dust features, which are present in some sources.

\subsection{The sample stars}

Here we describe the model fits to the spectra of
the individual stars,  which where analyzed in Paper I.
For a description of the
individual stars in this sample we refer to Paper I. From this
sample we rejected Roberts~22 and VY~2-2, because the ISO
satellite was unfortunately offset when observing these two
objects resulting in large flux jumps in the overall spectrum.
This made it impossible to derive temperature estimates of the
dust around these two stars. OH26.5+0.6 has also not been fitted,
because below 30~$\mu$m it has an absorption spectrum (Sylvester
et al. 1999), which could not be described with our simple model.

The main uncertainties in the model fits are due to uncertainties in
the continuum subtraction. 
This leads to errors in the temperature of the
order of 10~K and mass uncertainties of the order of a factor 2.
We note that for our modelling we completely rely on the laboratory data
input. This may result in systematic effects on our derived temperatures and
masses.

\subsubsection{IRAS09425-6050}
\label{sec:c_iras09}

The fit to the spectrum of IRAS09425-6040 is shown in 
Fig.~\ref{fig:iras09_fit}.
The model fit also produces a somewhat too strong 19.5 micron feature.
It should be noted that the full radiative transfer calculations of
Molster et al. (2001a) produces excellent fits to the 19.5~micron feature.
The broad feature at 11 $\mu$m is due to SiC. This very simple model
predicts no significant flux in the 10 micron complex due to
crystalline silicates, which is consistent with its absence in the ISO
spectrum.

The forsterite grains have a temperature of 85~K.
This temperature agrees with the temperature range presented in the detailed
radiative transfer model (Molster et al. 2001a). 
However, in contrast to the results presented 
here these detailed calculations predict that enstatite is much cooler than 
forsterite. As a result those models could not reproduce the relative strength
of the observed 28 and 40 micron complexes.
It also resulted in an unrealistically high mass for the enstatite.
Molster et al (2001a) argue that this might have to do with the not well known
absorptivity of crystalline enstatite.

\begin{figure}[t]
\centerline{\psfig{figure=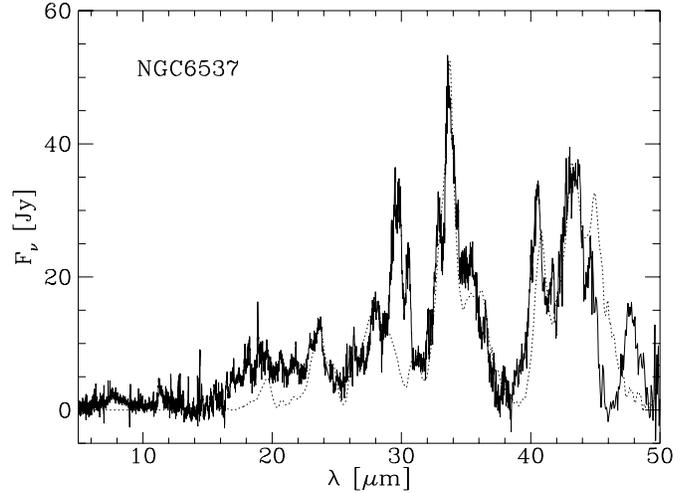,width=88mm,angle=270}}
\caption[]{\small A fit (dotted line) to the continuum subtracted spectrum (solid line) of NGC6537. $T_f = 75$~K and $T_e = 65$~K.
}
\label{fig:ngc6537_fit}
\end{figure}

\subsubsection{NGC6537}

The results for NGC6537 are shown in Fig.~\ref{fig:ngc6537_fit}.
The temperatures found for the forsterite (75~K)
and enstatite (60~K) in NGC6537 are among the lowest found in our sample.
Note that if an extra dust component
significantly contributes to the 40 micron complex, the temperature
of enstatite will be higher (and its mass lower)
than what has been determined here.

The spectral energy distribution of the complete spectrum is too broad to be 
fitted by a single temperature dust component. 

\begin{figure}[t]
\centerline{\psfig{figure=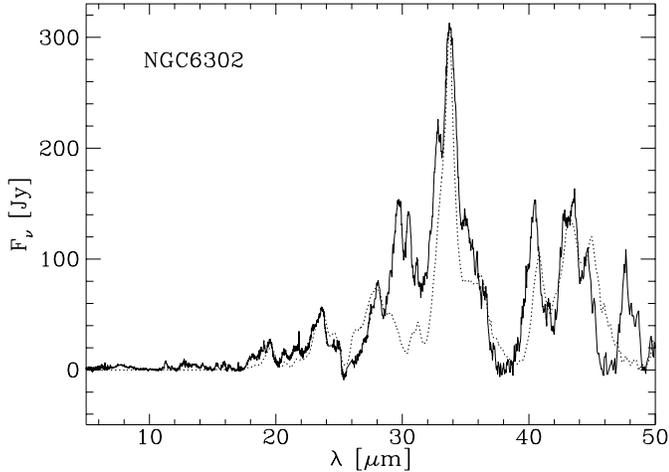,width=88mm,angle=270}}
\caption[]{\small A fit (dotted line) to the continuum subtracted spectrum (solid line) of NGC6302. $T_f = 65$~K and $T_e = 70$~K.
}
\label{fig:ngc6302_fit}
\end{figure}

\subsubsection{NGC6302}

The continuum subtracted spectrum of NGC 6302 and its good fit are
shown in Fig.~\ref{fig:ngc6302_fit}.

Molster et al. (2001b) used the same method as used in this paper, and 
therefore found the same temperatures. As for NGC6537, it was not possible 
to fit the spectral energy distribution with a single temperature dust 
component. Molster et al. (2001b) attribute the broad energy distribution to
the presence of a population of large grains, which mainly contribute to the 
long wavelength side. The presence of this population of large grains
is indicated by the gentle slope of the spectrum up to mm 
wavelengths (Hoare et al. 1992).

The temperature found for the enstatite and forsterite, respectively
65 and 70~K, are similar to the temperature of NGC6537, which in many aspects 
looks very similar to NGC6302. 
Kemper et al. (2001) assumed two temperature regimes: a cold one 
from 30 to 60~K, and a warm one from 100 to 118~K. Both components contain
forsterite and enstatite. Our results, giving a temperature somewhere 
in between those two regimes, is in agreement with theirs, although 
the exact comparison is somewhat difficult. 

\subsubsection{MWC922}

\begin{figure}[t]
\centerline{\psfig{figure=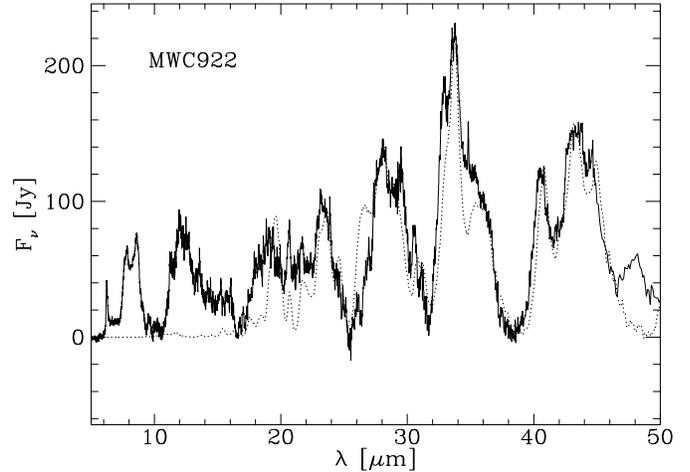,width=88mm,angle=270}}
\caption[]{\small A fit (dotted line) to the continuum subtracted spectrum (solid line) of MWC922. $T_f = 90$~K
and $T_e = 100$~K.
}
\label{fig:mwc922_fit}
\end{figure}

The fit to the continuum subtracted spectrum of MWC922 is one of
the best we have (see Fig~\ref{fig:mwc922_fit}). 
Especially the 40 micron complex is very well
reproduced by our model, indicating that the 50\% clino- and 50\%
ortho-enstatite are the right proportions for this object.
At $\lambda < 16 \mu$m the spectrum of MWC922
is dominated by PAH-features which
were not incorporated in the fitting procedure.

\subsubsection{AC~Her}

\begin{figure}[t]
\centerline{\psfig{figure=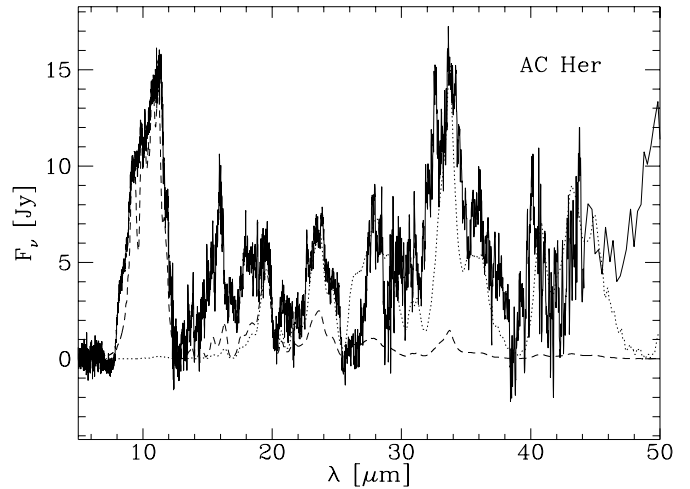,width=88mm,angle=270}}
\caption[]{\small A fit (dotted line) to the continuum subtracted spectrum (solid line) of AC Her.
$T_f = 100$~K and $T_e = 90$~K (dotted
line). The dashed line is a 700~K (for both forsterite and enstatite)
model fit.}
\label{fig:acher_fit}
\end{figure}

A model with cool dust fits the long wavelength part ($>20\mu$m)
of the AC Her spectrum
(dotted line in Fig.~\ref{fig:acher_fit}).
However, the short wavelength features indicate the presence of
a dust component with a much higher temperature.
The temperature of this material is not well constrained.
In Fig.~\ref{fig:acher_fit}
we show a fit of 700~K (dashed line in Fig.~\ref{fig:acher_fit}),
but a similar fit could be derived with a temperature several hundreds 
degrees Kelvin higher or lower. Therefore it 
is impossible to give a reliable mass estimate for this hot component.

In our modelling we only assumed a single temperature. Based on the
necessity of (at least) two different temperatures, the existence of a
temperature gradient seems more likely.
It is interesting to note that the overall
spectrum of AC Her is very similar to that of comet Hale Bopp (Molster et al. 1999a)
where we know that the dust is located in one place.
Temperature differences found in the grains around this comet must therefore
originate from the grain size differences. Small grains can
account for the high temperature dust emission, while bigger grains are
responsible for the low temperature dust emission.
Such a scenario might also be possible for AC Her, which would imply that the
dust might not have to be so close to the star as previously thought
(e.g. Alcolea \& Bujarrabal 1991).
Jura et al. (2000) found a disk like structure for this object, which
supports the above mentioned scenario.
A full radiative transfer model fit would be necessary to completely understand
the dust distribution around AC~Her, but that is beyond the scope of this 
paper.

\subsubsection{HD45677}

\begin{figure}[t]
\centerline{\psfig{figure=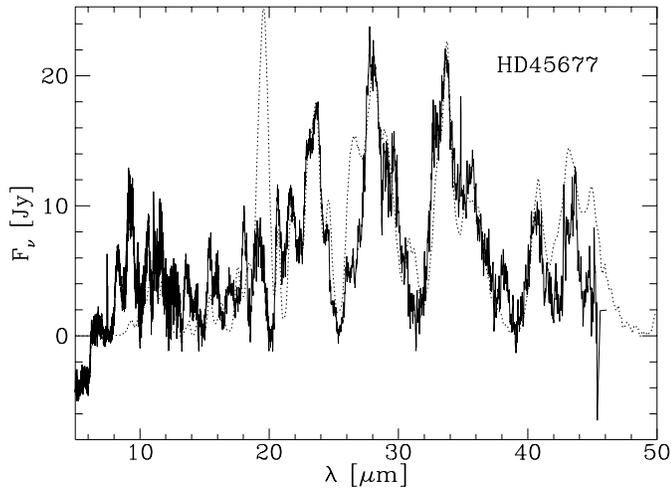,width=88mm,angle=270}}
\caption[]{\small A fit (dotted line) to the continuum and amorphous
silicate subtracted spectrum (solid line) of HD45677.
$T_f = 140$~K and $T_e = 140$~K.
}
\label{fig:hd45677_fit}
\end{figure}

From the continuum subtracted spectrum of HD45677 we first removed the
broad amorphous silicate features (Fig.~\ref{fig:hd45677_fit}). 
We cannot exclude that we also removed
part of the crystalline silicate features in the 18 micron complex
in this way. This does not influence our results since these are mainly based 
on the 23, 28, 33 and 40 micron complexes.
To fit the spectrum of HD45677 we ignored the strength
of the 19.5~micron feature, which is severely overestimated in our resulting 
fit. If we would have fitted the 19.5 and 40 micron features simultaneously,
the 28 micron complex would have been severely underestimated. 
Likewise, attempts to fit the 18 and 28 micron complex together will result 
in a severely overestimated 40 micron complex, and also the fits
to the 23 and 33 micron complexes will become worse.
It is unlikely that this discrepancy can be fully explained
by the subtraction of the amorphous silicates. 
Because this is not the only source with this problem, we leave this for
future research.

Malfait (1999) also studied this star. He modelled this object with
a radiative transfer code. HD45677 could only be modelled
with a 2 component dust shell, consisting of a hot shell, responsible
for the main part of the flux up to 20~$\mu$m and a cool component
which is the main contributor to the crystalline silicates features.
Due to the method we use here, our temperature estimate
is based on this cool component. Malfait finds a temperature between 250
and 50~K for this cool component. Unfortunately this is not
specified for the different components separately, so we
can only say that our temperature
estimates do agree with this temperature range.

The predicted strength of the crystalline silicate features in the 10 micron
complex is underestimated.
Since the strength of the amorphous silicate band at 10 $\mu$m is
uncertain, errors in the estimate of its contribution affect the
strength of the crystalline silicate bands at these wavelengths and
we did not attempt to fit the hot crystalline silicate compounds.

\subsubsection{89~Her}

\begin{figure}[t]
\centerline{\psfig{figure=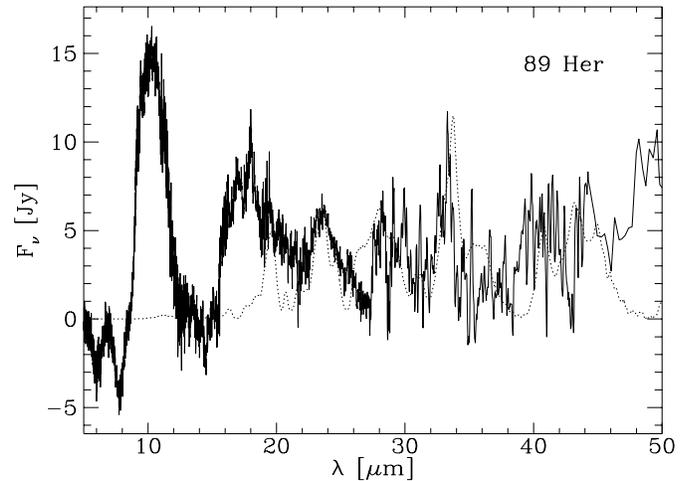,width=88mm,angle=270}}
\caption[]{\small A fit (dotted line) to the continuum subtracted spectrum (solid line) of 89~Her.
$T_f = 110$~K and $T_e = 100$~K.
}
\label{fig:89her_fit}
\end{figure}

Before we fitted the continuum subtracted spectrum of 89~Her, we first
subtracted a broad feature below the 26 to 45 $\mu$m region 
(Fig.~\ref{fig:89her_fit}). This
feature is also seen in HD44179 and probably AFGL~4106
and discussed in Paper II.

The continuum subtracted spectrum of 89 Her is quite noisy at the longer
wavelengths, which makes the fits not as well constrained as in other stars.
Also in this star warmer grains are necessary to explain the
crystalline silicate structure found on top of the amorphous silicate feature
in the 10 micron complex.
Again, problems in the separation of the crystalline and amorphous
silicates kept us from fitting this feature.
Based on the CO observations and the near-IR excess, it was argued
in Paper I, that there must be dust with different temperatures,
likely a temperature gradient, around 89~Her.

\subsubsection{MWC300}
\label{sec:c_mwc300}

\begin{figure}[t]
\centerline{\psfig{figure=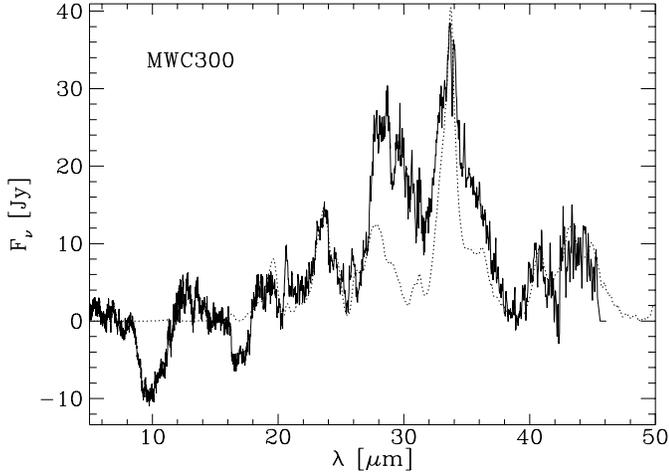,width=88mm,angle=270}}
\caption[]{\small A fit (dotted line) to the continuum subtracted spectrum (solid line) of MWC300.
$T_f = 90$~K and $T_e = 90$~K.
}
\label{fig:mwc300_fit}
\end{figure}

Although we have argued in this paper that the strength
of the 19.5 micron feature is difficult to model correctly,
we decided, because of the problems in the 28 micron complex
to constrain the enstatite by the 19.5 micron feature in MWC300
(see Fig.~\ref{fig:mwc300_fit}). If we would have fitted the
strength of the 28 micron complex, and ignored the 19.5 micron feature, we
would have derived a temperature of roughly 150~K, which is much
larger than the forsterite temperature. It would also predict
prominent features at the shorter wavelengths, which were not seen.
The use of the 19.5 micron feature to constrain enstatite
resulted in a similar temperature for the forsterite
and enstatite dust species. This result, together with the reasonable fit at 
the 40 micron complex makes us confident in our
approach for this star. We note however, that because of the problems
with the 28 micron feature the values for enstatite are poorly constrained.
The source of the extra flux in the 28 micron feature is unknown.

\subsubsection{HD44179}

\begin{figure}[t]
\centerline{\psfig{figure=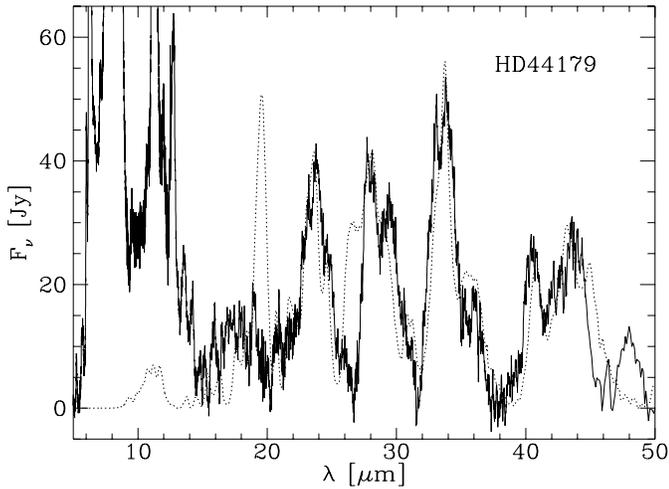,width=88mm,angle=270}}
\caption[]{\small A fit (dotted line) to the continuum subtracted spectrum (solid line) of HD44179. $T_f = 135$~K and $T_e = 135$~K.
Below 15 $\mu$m the spectrum is dominated by PAH features.}
\label{fig:rr_fit}
\end{figure}

As for 89 Her, we removed the very broad feature in the
26 to 45 $\mu$m range in the continuum subtracted spectrum of HD44179.
The 18 micron complex seems to contain
a contribution from amorphous silicates which was also removed.
The result can be found in Fig.~\ref{fig:rr_fit}.
It should be noted that a change in the subtraction of the broad 18 micron
amorphous silicate feature, whose properties are not well determined,
may change the derived temperature by more than the
typical fitting error of 10~K.

The derived continuum temperature of 120~K is somewhat uncertain due to
the complex nature of the source (Waters et al. 1998).
Since the short wavelength part of the continuum is formed by C-rich grains,
we based our fits on the long wavelength range. This will likely underestimate
the temperature of the continuum. Taken all these uncertainties into
account, the crystalline silicates may have an equal
or even lower temperature than the amorphous silicates, as is found
in the other stars.

\subsubsection{HD161796}

\begin{figure}[t]
\centerline{\psfig{figure=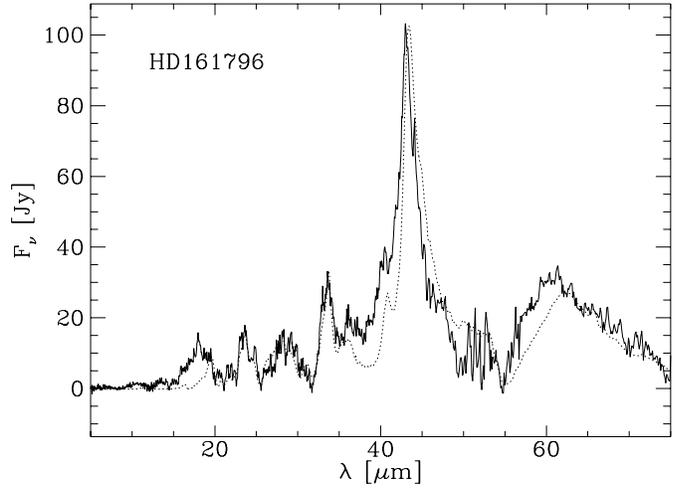,width=88mm,angle=270}}
\caption[]{\small A fit (dotted line) to the continuum subtracted spectrum 
(solid line) of HD161796. 
$T_f = 105$~K and $T_e = 80$~K.
We also included crystalline water ice with a temperature of 70~K}
\label{fig:hd161796_fit}
\end{figure}

The 40 micron complex in HD161796 is dominated by crystalline water ice.
In order to fit this spectrum (Fig.~\ref{fig:hd161796_fit}) we 
therefore added crystalline water ice to the spectrum (Smith et al. 1994). 
In general most features are reasonably well reproduced. 
The poor fit around 40 $\mu$m suggests that an underlying weak broad
component contributes. Possibly, this is amorphous water ice.

Within the errors, the crystalline and amorphous temperatures are the same.
The peak wavelength of the crystalline water ice feature in our model 
is slightly offset from our ISO spectrum. This may be a temperature effect 
(Smith et al. 1994), reflecting the sensitivity of the emissivities to 
temperature.

\subsubsection{HD179821}

\begin{figure}[t]
\centerline{\psfig{figure=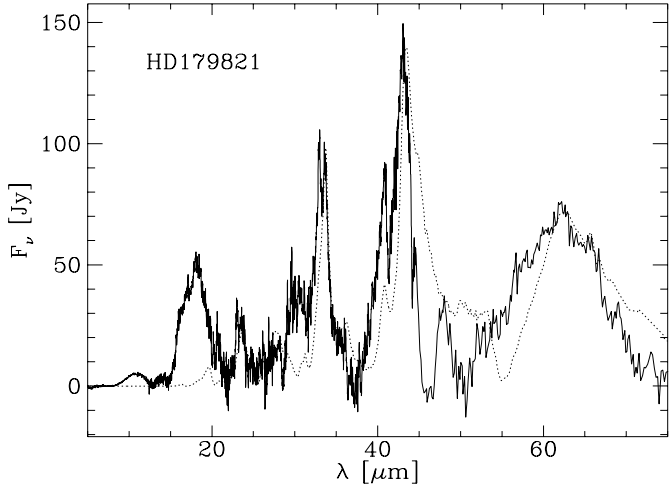,width=88mm,angle=270}}
\caption[]{\small A fit (dotted line) to the continuum subtracted spectrum (solid
line) of HD179821. $T_f = 75$~K and $T_e = 65$~K.
To fit the 40 and 60 micron complex, we included crystalline
water ice with a temperature of 45~K.
Note, that we made no attempt to fit the amorphous silicate bands in the 
continuum subtracted spectrum}
\label{fig:hd179821_fit}
\end{figure}

HD179821 shows prominent crystalline water ice features in the 40 and 60
micron complex. 
Therefore, we also added crystalline water ice in this model fit 
(Fig.~\ref{fig:hd179821_fit}). 
The crystalline silicate features are rather cold, 65~K for the enstatite and
75~K for the forsterite, therefore no detectable features
are expected in the 10 micron region.

\subsubsection{AFGL4106}

\begin{figure}[t]
\centerline{\psfig{figure=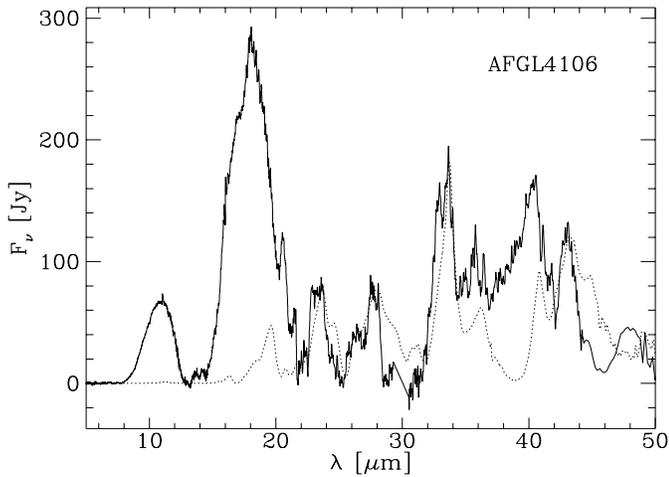,width=88mm,angle=270}}
\caption[]{\small A fit (dotted line) to the continuum subtracted spectrum (solid line) of AFGL4106.
$T_f = 100$~K and $T_e = 80$~K.
Note, that we made no attempt to fit the amorphous silicate bands in the 
continuum subtracted spectrum}
\label{fig:afgl4106_fit}
\end{figure}

For the fit to the continuum subtracted spectrum of AFGL4106
(Fig.~\ref{fig:afgl4106_fit}) we changed the ratio of clino versus ortho
enstatite to 1 : 3. Otherwise the 44.7 micron feature would have appeared too
strong in the model spectrum. This feature is located in the wing of the 43.0
micron feature. The 43.0 micron feature is not well reproduced by our simple 
model, resulting in an offset of the 44.7 micron feature. 

A broad feature between 30 and 45 $\mu$m seems present. It is not known
whether this is the same feature as in 89 Her and HD44179. It seems to peak at
a longer wavelength (38 $\mu$m) than in the other two sources.
Subtracting this feature, would increase both the temperature of
forsterite and enstatite.
The 40.5 $\mu$m feature is very strong, even if one removes this broad feature,
indicating that other dust species are present.
No attempt has been made to fit the substructure in the 18 micron band in view 
of the large and uncertain amorphous silicate contribution.
In this star no
detectable 10 micron structure is expected from the crystalline silicates,
which is in agreement with the observations.

A full radiative transfer calculation was made for this source
by Molster et al. (1999b).
The enstatite abundance derived by them was based on the (wrong) 
assumption that the 32.8 micron feature was due to ortho-enstatite,
and is therefore difficult to compare. 
This is related to their use of another datasets for forsterite, which has 
a different intrinsic ratio of the 23.7 and 33.6 micron features.
The result is a lower temperature and therefore higher mass for forsterite in
their calculations. Also, for the amorphous olivines they used a different 
dataset, however in this case it resulted in a higher temperature and 
therefore lower mass. The combination of those two effects explains the low
amorphous silicate to forsterite ratio found by them compared to this study.

\subsubsection{NML~Cyg}

We could not reliably fit enstatite to the spectrum of NML~Cyg since
a part of the 28 micron complex was missing, due to problems
with the data reduction (see Paper I). We were able to
fit forsterite to the continuum subtracted spectrum and found a
temperature of about 150~K (not shown).

\subsubsection{IRC+10420}

Also for IRC+10420 we only fitted the forsterite component which appeared
to be about 90~K (not shown).
Enstatite could not be fitted due to the lack of data in the 28 micron region.
We checked the 40 micron complex and, if enstatite would have been fitted,
an ortho-enstatite to clino-enstatite ratio of 2 : 1 would probably give the 
best fit to the strength of the 44.7 micron feature.

\section{Correlations}
\label{sec:corr}

From the modelling, we were able to derive values for the temperature and the
relative contributions to the total amount of dust mass by
forsterite and enstatite.
Correlating these values can give interesting insight in the
dust inventory as we will show in this section.
Since in Paper I and II it was found that there is a significant difference between
the spectra of disk and of outflow sources, we will separate them.
In all plots in this section the outflow sources are represented
by a circle while the disk sources are represented by diamonds.

\subsection{Temperature dependence of laboratory spectra: 
peak width and position}

\begin{figure}[t]
\centerline{\psfig{figure=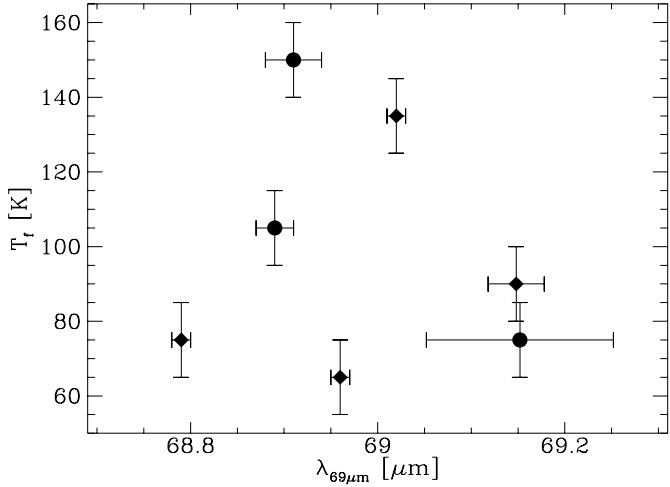,width=88mm,angle=270}}
\caption[]{\small The temperature of the forsterite grains versus the wavelength
position of the 69 micron band. The circles are outflow sources and
the diamonds are disk sources.}
\label{fig:T_69}
\end{figure}

In Sec~\ref{sec:comp} we showed a relation between the peak
position and FWHM of the 69 micron forsterite band and the temperature of the
forsterite grains.  Since we have now determined the temperature of the 
forsterite grains, we checked this relation in our data (Fig.~\ref{fig:T_69}).
There is no clear correlation between these two quantities.
A possible explanation for this scatter behaviour might be the fact that the
temperature of the forsterite
is determined from bands in the 20 to 40 $\mu$m range. The strength and width
of the 69 micron feature may be dominated by much cooler dust.
The predicted strength from our simple model fits, which is lower than
the strength in our ISO spectra (see e.g. HD179821), supports this statement.

\subsection{Other temperature trends}
\label{sec:Tempmass}

\begin{figure}[t]
\centerline{\psfig{figure=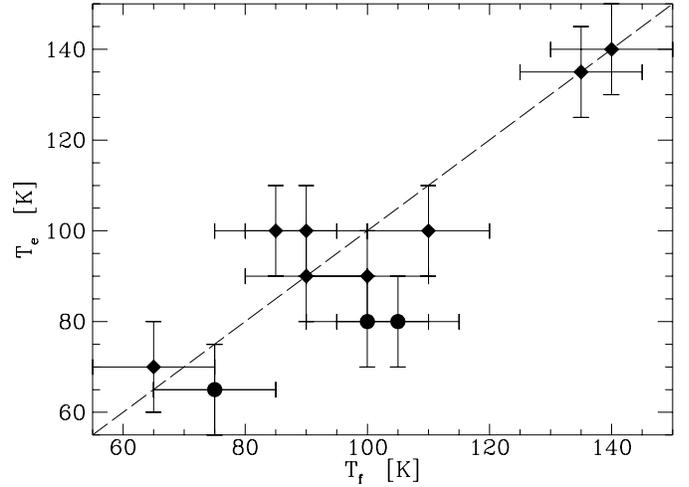,width=88mm,angle=270}}
\caption[]{\small The temperature of the forsterite grains versus the 
temperature of
the enstatite grains. The diamonds are the disk and the circles are the 
outflow sources. The dashed line represents equal temperatures for
the forsterite and enstatite grains. The errors are typically in the order of
10~K}
\label{fig:tempforsens}
\end{figure}

In Figure~\ref{fig:tempforsens} we compare the temperature of the
enstatite and forsterite, derived from our simple model fits. For the disk
sources the enstatite and forsterite grains seem to have an equal temperature,
while in the 3 outflow sources the forsterite seems warmer. 

\begin{figure}[t]
\centerline{\psfig{figure=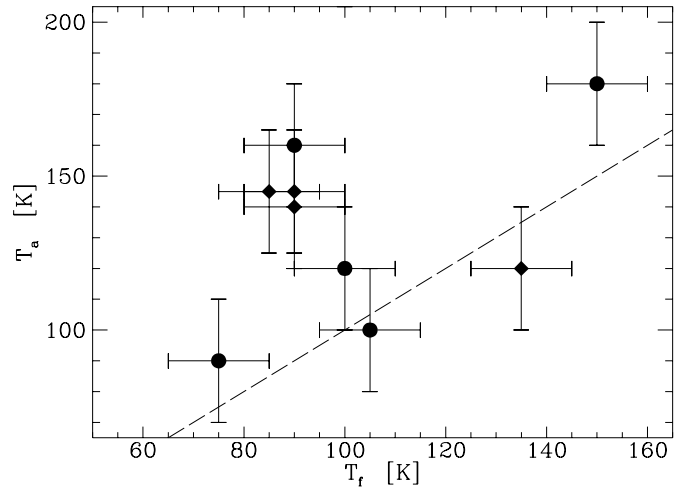,width=88mm,angle=270}}
\caption[]{\small The temperature of the forsterite grains versus the 
temperature of the amorphous silicate grains. The dashed line represents
equal temperatures for
the forsterite and amorphous silicate grains. The diamonds correspond to the
disk sources, while the circles correspond to the outflow sources.}
\label{fig:tempforsamorf}
\end{figure}

In Fig.~\ref{fig:tempforsamorf} we compare the forsterite temperature and 
the amorphous silicate temperature.
We first note that, in general, the crystalline forsterite
grains are colder than the underlying continuum consisting of
amorphous silicates. 
The difference in temperature between the amorphous and crystalline silicates
has probably to do with the difference in chemical structure.
The crystalline silicates are Mg-rich (see this paper and Paper II),
while the amorphous silicates must contain metals to explain their higher
near-infrared absorptivity.
Possibly the reaction of the Mg-rich crystalline silicates with gaseous iron
may proceed in these outflows at temperatures well below the
glass temperature leading simultaneously to amorphous and dirty silicates
(Tielens et al. 1998).

For some stars like IRC+10420, MWC922 and IRAS09425-6040 the difference in
temperature excludes the possibility of one grain population,
which is partially crystalline. The crystalline and amorphous silicates
must form two separate grain populations, which are not in thermal contact.
From this simple model we cannot say that they are also spatially
separated, but radiative transfer modelling (Molster et al. 1999b; 2001a) 
indicates that this is not necessary.
Finally, we note that there is no obvious separation between the disk and the
outflow sources. Because of the analogy between the different sources, we
expect that this grain segregation is valid for all sources in our sample.

\subsection{Abundance trends}
\label{sec:abund}

\begin{figure}[t]
\centerline{\psfig{figure=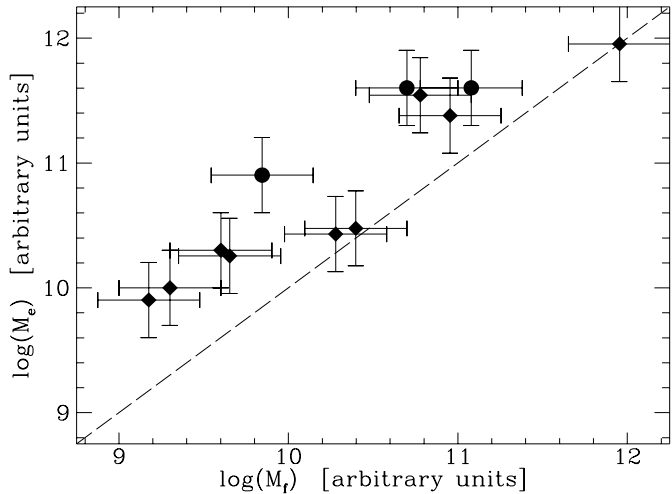,width=88mm,angle=270}}
\caption[]{\small The mass of the forsterite grains versus the mass of the enstatite
grains. We have assumed that they have the same size and shape distribution.
The dashed line represents equal masses for the forsterite and enstatite
grains. The outflow sources are marked as a circle while the disk sources are
marked with a diamond.}
\label{fig:massforsens}
\end{figure}

From our modelling we were also able to derive
an abundance ratio for enstatite and forsterite (Fig.~\ref{fig:massforsens}).
In all sources, except NGC6302, the
enstatite is more abundant than the forsterite, on average a factor 3--4.
Again there seems to be a difference between the disk and outflow sources. 
The outflow sources have, on average, a higher enstatite over forsterite 
ratio than the disk sources. 
We note however, that the number of outflow sources is low and more
data is required to confirm this trend.

If crystallization of the amorphous grains is a thermally driven
process, it is expected that warmer dust shells would show a higher
crystalline silicate fraction. 
This effect would be most clear in the disks sources, where
the dust stays at roughly the same place. In the outflow sources,
the temperature of the optically thin dust shell is more an indication of
the age of the dust. Cooler dust would be further out and therefore older,
in which case an increase of the crystallinity with a lowering of
the temperature may even be possible.
We investigated the crystalline over amorphous silicates
abundance ratio with the temperature of the amorphous silicates.
We excluded AC~Her, 89~Her and HD45677 from this analysis because the
continuum and feature temperatures derived likely do not reflect
particles with the same size distribution and/or location around
these objects, resulting in unrealistic mass ratios.
No clear relation could be found between the temperature of the amorphous
silicates versus the crystallinity, neither for the disk nor for the outflow
sources. The crystallization process is either a relic of the past or it is
not thermally driven.
This is not too surprising. Thermally driven crystallization requires
temperatures of $\approx 1000$K (Hallenbeck \& Nuth, 1998), which is well 
above the temperature of either the crystalline or amorphous silicates
in our sources.

\subsection{Crystallinity versus stellar flux ratio}
\label{sec:cryst_fir}

It has been suggested by Waters et al. (1999) that the enstatite
over forsterite ratio increases with luminosity of the star.
They based their conclusions on the ratio of the
32.97 and 33.6 micron bands. They attributed the former band to enstatite,
and the latter one to forsterite. Unfortunately, the 32.97 micron band
is strongly affected by instrumental effects
(see Paper II), which are most prominent in the
brightest sources. These are usually also the intrinsic most
luminous sources, such as IRC+10420 and NML~Cyg.
Since it is likely that in high flux sources grain formation
will result in different ratios of the dust species we still decided
to investigated this scenario. We could find no trend for the stars in our 
sample, which had a reliable distance (and thus flux) estimate.

\section{Discussion}

Most features can be explained with forsterite and enstatite. However
laboratory spectra of these species often produce too broad bands. Most
laboratory spectra were measured at room temperature, while the dust species
around our stars are in the order of 80~K. Lowering the temperature can narrow
the dust features, and this might be the
natural solution for the width problem.  It can also solve another problem.
The 69.0 micron feature has been attributed to forsterite. It is very sensitive
to the Fe/(Mg+Fe) ratio in olivine, and one of
the key features for the determination of the Fe content in crystalline
silicates. However in most lab spectra it peaks at 69.7~$\mu$m. Recent 
laboratory measurements show that this feature shifts to shorter wavelengths 
when the sample is cooled and will be at $\approx 69.1~\mu$m at 77~K and even 
at $\approx 68.8 \mu$m when measured at 4~K (Bowey et al. 2000). 
At the same time, the feature narrows (cf. Fig~\ref{fig:69T}). So it seems
that the temperature dependence is able to solve also this problem.

From the correlation study in this chapter a clear difference
between the disk and the outflow sources emerges. The differences
found in Paper I and Paper II were based on the shape and strength of the
different solid state features in the spectrum. These differences
can now be traced back to differences in temperature and chemical
composition of the circumstellar dust. The outflow sources seem
more abundant in cold enstatite grains than the disk sources,
although more observations would help to quantify this better.
There are also indications that in the high luminosity outflow
sources the ortho- over clino-enstatite abundance ratio is larger
than unity (see AFGL4106 and IRC+10420), while in the other
(disk) sources nice fits were obtained with a ratio of unity.
Because of the large crystalline water-ice component in the 40
micron complex of HD161796, it is difficult to quantify its
ortho- over clino-enstatite ratio. Therefore we cannot exclude
that this difference we see is related to the more massive nature
of the central stars of AFGL4106 and IRC+10420 rather than the
disk-outflow character.

The difference between the disk and the outflow sources might be related
to differences in conditions during the condensation of the grains out
of the hot
gas, and/or to differences in the conditions since the formation of
the dust particles. Smyth (1974) found that clino-enstatite
slowely inverts to ortho-enstatite between 920 and 1220~K. Dust particles
will stay longer within this temperature range in the case of massive stars
than for low mass stars. This might explain the overabundance of
ortho-enstatite in IRC+10420 and AFGL4106 with respect to the other stars.
Another possible difference during the formation
of these grains could be the amount of radiation pressure exerted on the
particles. It is expected to be lower
in the disk sources, otherwise the disk is likely to be
blown away. Also the presence of a companion -most if not all disk sources
are a binary system- might influence the dust forming process.
If, on the other hand, we assume that the initial (forming) conditions
are similar for the disk and outflow sources, than the differences in the
conditions after formation must be dominant. Time is an obvious
differences. Also a relation with grain coagulation -the disks contain
large grains- cannot completely be ruled out.

We note that clino-enstatite is often found
in meteorites on earth. The process responsible for the overabundance
of clino-enstatite in meteorites in the disks of young stars might be the same
as the one in the disks around evolved stars.
We also want to note that in chondrites or primitive meteorites, i.e
the less processed ones, clino-pyroxenes are less abundant than the 
ortho-pyroxenes.
This similarity between the disks around young and around evolved stars, is
new evidence to the hypothesis that the
circumstances and processes in disks around young stars are very similar
to these in disks around evolved stars, although
their origin is quite different.

The equal temperature of enstatite and forsterite in the disk 
sources, is in principle compatible with the assumption that enstatite and 
forsterite are present as a composite grain in the disk sources. 
However, it is equally well possible that their individual equilibrium 
temperatures are similar, in which case they can still be 
individual grains. 
New uninterupted laboratory measurements of the optical properties of both 
materials, especially in the wavelength range were they absorb the stellar 
light and emit the thermal radiation, would allow us to make these kind of 
calculations and settle this point.
The different temperatures found for forsterite and enstatite in the outflow
sources suggests two separate grain populations in these environments.

\section{Conclusions}

We can summarize the main results of this study as follows:
\begin{itemize}
\item[1] The ISO spectra could be reasonably well fitted with laboratory
spectra of forsterite and enstatite.
\item[2] The models underestimate the flux at 18, 29.6, 30.6, 48 $\mu$m
and sometimes at 40.5 $\mu$m,
which is an indication for the presence of (an)other dust component(s).
No convincing identification could be made yet. Diopside does have features at
most of these wavelengths, but also strong features at others which are
weak or absent in the ISO-spectra.
\item[3] The 19.5 micron feature is often overestimated by our model spectra.
No explanation is yet known for this phenomenon, but it should be noted
that in the full radiative transfer modelling it appeared to be much less
of a problem. This might indicate that optical depth effects play a role.
Also the calculation of the absorption coefficients from the
constants instead of the absorption coefficients from laboratory particles
might lead to differences.
\item[4] The band width of the laboratory data is larger than in 
our ISO spectra. This difference is likely a temperature effect, and
might be used as an independent temperature indicator. Especially the
69.0 micron band is very suitable for this analysis.
\item[5] The temperature of the forsterite and enstatite grains
are (almost) similar for the disk sources, while the forsterite
is slightly warmer in the outflow sources. This would imply that
the forsterite and/or enstatite grains differ slightly in
the disk and outflow sources. It is not clear whether this
difference is due to a different formation process, or due to a different
dust process history after the grain formation.
Since this trend is
only based on 3 sources more data is required to confirm the difference 
between the dust and outflow sources.
\item[6] The crystalline silicates are colder than the amorphous silicates.
This is probably due to the difference in chemical composition.
No Fe is present in the crystalline silicates, while in the amorphous
silicates it is expected to explain the higher absorptivity. This difference
in temperature also implies that the crystalline and amorphous grains are two
distinct grain populations.
\item[7] Enstatite is on average a factor 3--4 more abundant than forsterite
in our sources. There are
indications that the enstatite over forsterite ratio in the outflow sources
is higher than in the disk sources.
\item[8] In the low luminosity sources the spectra were well fitted with
an equal amount of ortho- and clino-enstatite, while in the two high
luminosity sources more ortho-enstatite seems to be present. 
\item[9] No correlation could be found between the crystallinity and the
temperature of the dust. Also the luminosity of the stars does not seem to be
correlated with the enstatite over forsterite ratio.
\item[10] These simple model fits already give a good insight in the dust
around our stars. In Paper I the shape of the features naturally separated the
disk and outflow sources. In this study the differences between these two
categories became again evident.
\end{itemize}

\vspace{1.0cm}
\noindent{\bf Acknowledgements.}\\
We greatly thank Janet Bowey for providing her laboratory data.
FJM wants to acknowledge the support from NWO under grant 781-71-052 and under
the Talent fellowship programm. LBFMW acknowledges financial support from an 
NWO `Pionier' grant.


\begin{thebibliography}{}

\bibitem{}
Alcolea J. and Bujarrabal V., 1991, A\&A 245, 499
\bibitem{}
Bohren C.F., Huffman D.R., 1983, Absorption and scattering of light by small 
particles, John Wiley and Sons Inc., New York
\bibitem{}
Bowey J.E., Lee C., Tucker C. et al., 2000, in
ISO beyond the peak, ed A. Salama, ESA SP-456, 339
\bibitem{}{}
Bradley J.P., Brownlee D.E. and Veblen D.R., 1983, Nature 301, 473
\bibitem{}
Chihara H., Koike C., Tsuchiyama A., 2001 PASJ 53, 243
\bibitem{}
Crovisier J., Leech K., Bockelee-Morvan D., et al., 1997, Science 275, 1904
\bibitem{}
Cohen M., Barlow M.J., Sylvester R.J. et al., 1999, ApJ 513, L135
\bibitem{}
Hallenbeck S. and Nuth J., 1998, A\&SS 255,427
\bibitem{}
Hoare M., Roche P.F., Clegg R.E.S., 1992, MNRAS 258, 257
\bibitem{}
J\"{a}ger C., Molster F.J., Dorschner J. et al., 1998, A\&A 339, 904
(JMD)
\bibitem{}
Jura M., Chen C. and Werner M.W., 2000, ApJ 541, 264
\bibitem{}
Kemper F., J\"ager C., Waters L.B.F.M. et al. 2001, submitted to Nature
\bibitem{}
Koike C., Shibai H. and Tuchiyama A., 1993, MNRAS 264, 654
\bibitem{}
Koike C. and Shibai H., 1994, MNRAS 269, 1011
\bibitem{}
Koike C. and Shibai H., 1998, ISAS report no. 671
\bibitem{}
Koike C., Tsuchiyama A., Shibai H., et al., 2000a, A\&A 363, 1115
\bibitem{}
Koike C., Chihara H., Tsuchiyama A. et al., 2000b, in 
Proceedings of the 33rd ISAS Lunar and Planetary Symposium, p 95 
\bibitem{}
Malfait K, Waelkens C., Waters L.B.F.M., 1998, A\&A 332, L25
\bibitem{}
Malfait K, 1999, {\em PhD thesis, Catholic University Leuven}
\bibitem{}
Mennella V., Brucato J.R., Colangeli L. et al., 1998, ApJ 496, 1058
\bibitem{}
Molster F.J., Yamamura I., Waters L.B.F.M., et al., 1999a, Nature 401, 563
\bibitem{}
Molster F.J., Waters L.B.F.M., Trams N. et al., 1999b, A\&A 350, 163
\bibitem{}
Molster F.J., Yamamura I., Waters L.B.F.M., et al., 2001a, A\&A 366, 923
\bibitem{}
Molster F.J., Lim T.L., Sylvester R.J. et al., 2001b, A\&A 372, 165
\bibitem{}
Molster F.J., Waters L.B.F.M., Tielens A.G.G.M., and Barlow M.J., 2001c, submitted to A\&A
(Paper I)
\bibitem{}
Molster F.J., Waters L.B.F.M., and Tielens A.G.G.M., 2001d, submitted to A\&A
(Paper II)
\bibitem{}
Ossenkopf V., Henning Th. and Mathis J.S., 1992, A\&A 261, 567
\bibitem{}
Servoin J.L., Piriou B., 1973, Phys. Stat. Sol. (B) 55, 677
\bibitem{}
Smith R.G., Robinson G., Hyland A.R., and Carpenter G.L., 1994, MNRAS 271, 481
\bibitem{}
Smyth J.R., 1974, American Mineralogist 59, 345
\bibitem{}
Sylvester R.J., Kemper F., Barlow M.J. et al., 1999 A\&A 352, 587
\bibitem{}
Tielens A.G.G.M., Waters L.B.F.M., Molster F.J. and Justtanont K., 1998, ApSS
255, 415
\bibitem{}
Voors R.H.M., 1999, {\em PhD thesis}, University of Utrecht
\bibitem{}
Waelkens C., Waters L.B.F.M., de Graauw M.S., et al., 1996, A\&A 315, L245
\bibitem{}
Waters L.B.F.M., Molster F.J., de Jong T. et al., 1996, A\&A 315, L361
\bibitem{}
Waters L.B.F.M., Waelkens C., van Winckel H. et al, 1998, Nature 391, 868
\bibitem{}
Waters L.B.F.M., Molster F.J., Waelkens C., 1999
IAU conference 191: ``Asymptotic Giant Branch Stars'',
eds. T. Le Bertre, A. L\`{e}bre and C. Waelkens, 209


\end{thebibliography}
\end{document}